# QED Plasma at High Temperature

Samina S. Masood[1]

Department of Physical and Applied Sciences, University of Houston
Clear Lake, Houston TX 77058

**ABSTRACT**

We demonstrate that the early universe behaved as a relativistic QED (Quantum Electrodynamics) plasma around the nucleosynthesis time while the temperature of the universe was below the neutrino decoupling temperature in the early universe. QED coupling constant becomes a temperature dependent parameter due to the radiative corrections to vacuum polarization in the early universe at nucleosynthesis temperature. Renormalization scheme of QED is used to calculate the effective parameters of relativistic plasma in the early universe. Renormalization constants of QED serve as effective parameters of the theory and are used to determine the behavior of matter. We explicitly compute the parameters of QED plasma such as Debye length and the plasma frequency as a function of temperature. Light is slowed down and trapped due to bending in such a medium and the frequency of electromagnetic radiation becomes a function of temperature as well.

1. INTRODUCTION

The QED coupling parameter alpha is constant in vacuum and does not depends on temperature because the electromagnetic waves are purely transverse in nature and the photon is massless in vacuum. It does not interact with the medium and the longitudinal component has to be zero to satisfy the requirement of gauge invariance of QED. When there were enough fermions in the universe, photons interact with the medium due to the vacuum polarization and acquired a dynamically generated mass because of its interaction with the fermions. QED coupling parameter alpha becomes a function of temperatures due to its interaction with the medium in the presence of the longitudinal component of the vacuum polarization tensor. Therefore, the QED coupling constant was very large in the presence of enough fermions in the early universe at temperatures greater than the electron mass. Above the neutrino decoupling temperatures, electrons were interacting with photons as well as neutrinos in the medium and QED was not able to fully describe this interaction. However, QED works perfectly fine below the neutrino decoupling temperature.

When the electromagnetic waves propagated in the early universe with extremely high temperatures the QED coupling depends on the temperature as the coupling of the propagating particles with the medium is affected by temperature. Renormalization scheme is used to compute vacuum polarization in such a medium in the real-time formalism [1, 2] and the effect of the interaction of radiation with fermions of the medium. Renormalization is a process of removal of singularities in gauge theories. Perturbation theory is needed to compute radiative corrections to higher order processes in a medium. Order by order cancellation of singularities is required by KLN (Kinoshita-Lee and Nauenberg) theorem [3, 4] to assure the finiteness of a theory at all orders of perturbative expansion. Interaction of electromagnetic radiation with the fermions of the medium through vacuum polarizations modifies the QED coupling, which when is sufficiently increased the phase transition can take place in the medium and the ideal gas approximation will not be applicable. Interaction with the medium induce temperature dependence to the physically measurable parameters [5-15] of the theory which continuously grow with temperature. It has already been shown that the QED coupling of electromagnetically interacting medium is associated with the dynamically generated plasma screening mass [3, 7, 13, 16] of photon and affects the propagation of photons in this medium. The phase of such a medium is determined by computing the plasma frequency and Debye shielding length as a function of temperature which correspond to the relativistic plasma in the early universe.

---

[1] Electronic address: masood@uhcl.edu

This paper is organized as follows: Section 2 briefly describes the calculational scheme of real-time formalism. This scheme is used to analytically calculate the vacuum polarization tensor of QED in Section 3 and the parameters of the relativistic plasma in Section 4. Results of Section 3 and 4 are applied to understand the propagation of particles in the early universe in Section 5.

## 2. CALCULATIONAL SCHEME

Thermal effects are included simply by replacing vacuum propagators by the modified propagators of a heatbath. Imaginary time formalism works in Euclidean space and time is taken as an imaginary coordinate and so is the energy. Renormalization scheme of QED in real-time formalism allows an order-by-order cancellation of singularities which cannot be done in Euclidean space. The corresponding real-time formalism in Minkowski space restore [2] the gauge invariance of QED with thermal propagator with the appropriate choice of the rest frame of the heatbath. The photon propagator of vacuum theory is modified as:

$$\frac{1}{k^2} \rightarrow \frac{1}{k^2} - 2\pi i \delta(k^2) n_B(k), \qquad (1)$$

and the photon distribution function of photons is given as:

$$n_B(k) = \left[\frac{1}{e^{\beta k_0} - 1}\right]. \qquad (1a)$$

Where β is the inverse temperature, i.e; 1/T and $k_0$ is the energy of the propagating photon. The corresponding fermion propagators is replaced as:

$$\frac{1}{p^2 - m^2} \rightarrow \left[\frac{1}{p^2 - m^2} + \Gamma_F(p)\right] \qquad (2)$$

And the fermion distribution function is expressed as

$$n_F(p) = \left[\frac{1}{e^{\beta(p_0)} + 1}\right] \qquad (2b)$$

With $p_0$ the energy of electron.

It has been shown [5, 8] that renormalization of QED in a hot medium of the early universe yields the self-mass (or self-energy) of electron as

$$\frac{\delta m}{m} \approx \frac{\alpha \pi T^2}{3m^2}\left[1 - \frac{6}{\pi^2}c(m\beta)\right] + \frac{2\alpha}{\pi}\frac{T}{m}a(m\beta) - \frac{3\alpha}{\pi}b(m\beta). \quad (3a)$$

With the wavefunction renormalization constant as

$$Z_2^{-1}(\beta) = Z_2^{-1}(T=0) - \frac{2\alpha}{\pi}\int_0^\infty \frac{dk}{k}n_B(k) - \frac{5\alpha}{\pi}b(m\beta)$$
$$+ \frac{\alpha T^2}{\pi v E^2}\ln\frac{1+v}{1-v}\left\{\frac{\pi^2}{6} + m\beta a(m\beta,\mu) - c(m\beta)\right\}, \quad (3b)$$

The electron charge renormalization constant [8] is:

$$Z_3 \cong 1 + \frac{2e^2}{\pi^2}\left[\frac{ma(m\beta)}{\beta} - \frac{c(m\beta)}{\beta^2} + \frac{1}{4}\left(m^2 + \frac{1}{3}\omega^2\right)b(m\beta)\right]. \quad (3c)$$

Where Masood's functions $a_i(m\beta)$ are given as:

$$a(m\beta) = \ln(1 + e^{-m\beta}), \quad (4a)$$

$$b(m\beta) = \sum_{n=1}^{\infty}(-1)^n \text{Ei}(-nm\beta), \quad (4b)$$

$$c(m\beta) = \sum_{n=1}^{\infty}(-1)^n \frac{e^{-nm\beta}}{n^2}, \quad (4c)$$

$Z_3$ in Eq. (3c) corresponds to the QED coupling constant $\alpha$ which is related to the charge e through the relation $\alpha = e^2/4\pi$ in the natural units.

### 3. Vacuum Polarization Tensor

The vacuum polarization tensor $\Pi_{\mu\nu}$ is a 4×4 matrix which describe the propagation of light in 4-dimensional space. Polarization of electromagnetic waves in 4-dimensional space and explains the propagation of light in 3-dimensional space with time. Transversality of light is associated with the zero mass of photon as a gauge requirement and is related to the absence of interaction of light with the medium and leads to the absence of its longitudinal component. However, in its general form, $\Pi_{\mu\nu}$ is expressed in a 4-dimensional space as

$$\Pi_{\mu\nu}(K,\mu) = P_{\mu\nu}\Pi_T(K,\mu) + Q_{\mu\nu}\pi\Pi_L(K,\mu) \quad (5a)$$

$\Pi_L(K,\mu)$ and $\Pi_T(K,\mu)$ correspond to the longitudinal and transverse components of the vacuum polarization tensor respectively. In an interacting fluid, the photon acquires dynamically generated mass which adds up a nonzero longitudinal component and the photons with nonzero screening mass are not quanta of a transverse wave and has the ability to interact with the medium. Particle properties of photons help to understand the electromagnetic properties of such a medium. The vacuum polarization tensor is expressed in terms of longitudinal and transverse components in Eq. (5b).

$$\Pi_{\mu\nu} = \begin{pmatrix} -\frac{k^2}{K^2}\pi_L & -\frac{i\omega k_1}{K^2}\pi_L & -\frac{i\omega k_2}{K^2}\pi_L & -\frac{i\omega k_3}{K^2}\pi_L \\ -\frac{i\omega k_1}{K^2}\pi_L & \left(-1-\frac{k_1^2}{k^2}\right)\pi_T + \left(\frac{\omega^2 k_1^2}{k^2 K^2}\right)\pi_L & \left(-\frac{k_1 k_2}{k^2}\right)\pi_T + \left(\frac{\omega^2 k_1 k_2}{k^2 K^2}\right)\pi_L & \left(-\frac{k_1 k_3}{k^2}\right)\pi_T + \left(\frac{\omega^2 k_1 k_3}{k^2 K^2}\right)\pi_L \\ -\frac{i\omega k_2}{K^2}\pi_L & \left(-\frac{k_1 k_2}{k^2}\right)\pi_T + \left(\frac{\omega^2 k_1 k_2}{k^2 K^2}\right)\pi_L & \left(-1-\frac{k_2^2}{k^2}\right)\pi_T + \left(\frac{\omega^2 k_2^2}{k^2 K^2}\right)\pi_L & \left(-\frac{k_2 k_3}{k^2}\right)\pi_T + \left(\frac{\omega^2 k_2 k_3}{k^2 K^2}\right)\pi_L \\ -\frac{i\omega k_3}{K^2}\pi_L & \left(-\frac{k_1 k_3}{k^2}\right)\pi_T + \left(\frac{\omega^2 k_1 k_3}{k^2 K^2}\right)\pi_L & \left(-\frac{k_2 k_3}{k^2}\right)\pi_T + \left(\frac{\omega^2 k_2 k_3}{k^2 K^2}\right)\pi_L & \left(-1-\frac{k_3^2}{k^2}\right)\pi_T + \left(\frac{\omega^2 k_3^2}{k^2 K^2}\right)\pi_L \end{pmatrix}$$

..................... (5b)

It shows that how the longitudinal component let the wave grow or die with time at the same point with the increase or decrease in the longitudinal component. $\Pi_{0i}$ or $\Pi_{i0}$ and even $\Pi_{00}$ components of the vacuum polarization vanish if the longitudinal component vanish. Energy change associated with the change in the longitudinal component indicates that the frequency and wavelength of electromagnetic signal will change with time depending on the change in $\Pi_L$ and $\Pi_T$ and it will affect the velocity of propagation as well. Other components of the vacuum polarization tensor are also affected with nonzero $\Pi_L$ indicating the change in distribution of signal in space and the modification in the polarization properties with temperature. Vacuum polarization tensor $\Pi_{\mu\nu}$ plays a key role in computing the electromagnetic properties of the medium itself and leads to the calculation of plasma generating mass and then the Debye shielding length indicating the phase change in to the relativistic plasmas. In addition to that, it can easily be seen if $\Pi_L=0$, the transverse wave will look like Eq. (6a)

$$P_{\mu\nu} = \begin{pmatrix} 0 & 0 & 0 & 0 \\ 0 & -1-\frac{k_1^2}{k^2} & -\frac{k_1 k_2}{k^2} & -\frac{k_1 k_3}{k^2} \\ 0 & -\frac{k_1 k_2}{k^2} & -1-\frac{k_2^2}{k^2} & -\frac{k_2 k_3}{k^2} \\ 0 & -\frac{k_1 k_3}{k^2} & -\frac{k_2 k_3}{k^2} & -1-\frac{k_3^2}{k^2} \end{pmatrix} \qquad (6a)$$

And if the electromagnetic signal were just the longitudinal with $\Pi_T=0$, it will look like Eq. (6b)

$$Q_{\mu\nu} = \begin{pmatrix} -\frac{k^2}{K^2} & -\frac{i\omega k_1}{K^2} & -\frac{i\omega k_2}{K^2} & -\frac{i\omega k_3}{K^2} \\ -\frac{i\omega k_1}{K^2} & \frac{\omega^2 k_1^2}{k^2 K^2} & \frac{\omega^2 k_1 k_2}{k^2 K^2} & \frac{\omega^2 k_1 k_3}{k^2 K^2} \\ -\frac{i\omega k_2}{K^2} & \frac{\omega^2 k_1 k_2}{k^2 K^2} & \frac{\omega^2 k_2^2}{k^2 K^2} & \frac{\omega^2 k_2 k_3}{k^2 K^2} \\ -\frac{i\omega k_3}{K^2} & \frac{\omega^2 k_1 k_3}{k^2 K^2} & \frac{\omega^2 k_2 k_3}{k^2 K^2} & \frac{\omega^2 k_3^2}{k^2 K^2} \end{pmatrix} \qquad (6b)$$

In the non-interacting fluids $\Pi_L=0$ and $\Pi_{\mu\nu}(K,\mu) = P_{\mu\nu}\Pi_T(K,\mu)$. Therefore nonzero value of $\Pi_L$ is a measure of deviation from the transverse behavior in an interacting fluid. If it is ignorable or sufficiently small, the fluid will affect the speed and have a small longitudinal component whereas it will have a sizeable effect for larger values of $\Pi_L$. When the neutrino decoupling takes place, enough neutrinos are present in the medium to interact with the virtual electrons which are produced during the vacuum polarization. Interaction of the neutrinos with these electrons give real contribution and the properties of such a medium are significantly modified as photon generates electrons which couple with neutrinos weakly and lead to electron-neutrino scattering processes. Interaction of virtual electrons with the magnetic field (or the photon in regular fluids or plasmon in interacting fluids) also contribute to the magnetic moment of neutrino [17-20] which has already been shown to vary with temperature and may also related to the refractive energy of the medium.

The longitudinal and transverse components of the vacuum polarization tensor $\Pi_L(K,\mu)$ and $\Pi_T(K,\mu)$ respectively are evaluated [5, 8] as

$$\Pi_L \cong \frac{4e^2}{\pi^2}(1-\frac{\omega^2}{k^2})[(1-\frac{\omega}{2k}\ln\frac{\omega+k}{\omega-k})(\frac{ma(m\beta)}{\beta}-\frac{c(m\beta)}{\beta^2})$$
$$+\frac{1}{4}(2m^2-\omega^2+\frac{11k^2+37\omega^2}{72})b(m\beta)], \qquad (7a)$$

and

$$\Pi_T \cong \frac{2e^2}{\pi^2}[\{\frac{\omega^2}{k^2}+(1-\frac{\omega^2}{k^2})\frac{\omega}{2k}\ln\frac{\omega+k}{\omega-k}\}(\frac{ma(m\beta)}{\beta}-\frac{c(m\beta)}{\beta^2})$$
$$+\frac{1}{8}(2m^2+\omega^2+\frac{107\omega^2+131k^2}{72})b(m\beta)]. \qquad (7b)$$

This information of temperature dependence of the longitudinal and transverse components of vacuum polarization tensor help to explicitly understand the electromagnetic properties of a medium at extremely high temperatures as well as the phase transition from an ideal gas to interacting fluid and relativistic plasma with the rise of temperature.

### 4. PARAMETRS OF QED PLASMA

The effective parameters of QED using the renormalization scheme can be described as the effective parameters in a QED plasma. Photon acquire plasma screening mass and affect the coupling constant which changes the electromagnetic properties of the medium. It is shown [1, 5, 8] that the photons in this medium develop a plasma screening mass which can be obtained from the longitudinal and transverse component of the vacuum polarization tensor $\Pi_L(0,k)$ and $\Pi_T(k,k)$ where $K^2 = \omega^2 - k^2 = 0$ in vacuum [2] equating $\omega^2 = k^2$. This transversality condition changes with temperature lead to the phase transition in a medium.

Longitudinal and transverse components ($\Pi_L$ and $\Pi_T$, respectively) of vacuum polarization tensor $\Pi_{\mu\nu}$ play a crucial role in the determination of the electromagnetic properties of such a medium. Parameters such as propagation speed $v_{prop}$, refractive index $i_r$, dielectric constant, electric permittivity $\varepsilon(K)$, magnetic permeability $\mu(K)$, and the magnetic moment $\mu_a$ of electromagnetically interacting particles propagating in the medium can be evaluated in detail to determine the behavior of the medium. All of these parameters are expressed in terms of $\Pi_L$ and $\Pi_T$ [4, 8] as:

$$\varepsilon(K) = 1 - \frac{\Pi_L}{K^2}, \qquad (8a)$$

$$\frac{1}{\mu(K)} = 1 + \frac{K^2 \Pi_T - \omega^2 \Pi_L}{k^2 K^2}, \tag{8b}$$

And [16]

$$\varepsilon(K) = 1 + \chi_e \tag{8c}$$

$$\mu(K) = 1 + \chi_m \tag{8d}$$

Whereas, $\chi_e$ and $\chi_m$ give the dielectric constant and magnetization of a medium at a given temperature, respectively. The relative change in the refractive index n with K can be evaluated as:

$$n(K) = \sqrt{\mu \varepsilon}$$

And the inverse of magnetic permeability correspond to the magnetic reluctance of the medium. The longitudinal and transverse components can be evaluated from the vacuum polarization tensor directly by using appropriate limits of photon frequency $\omega$ and the wavenumber k. The propagation of particles in a medium is affected by the interaction of the particles with a medium. In free space, the speed of light is expressed as

$$c = \sqrt{\frac{1}{\varepsilon_0(K) \mu_0(K)}}$$

For purely transverse signals the refractive index is related to the opacity of a medium as the light can be blocked and delayed at high temperatures. The larger temperature may even trap light due to bending. However the longitudinal frequency and the wavelength is zero due to the absence of interaction with the medium. At high temperatures of the early universe, excessive production of fermions lead to the sizeable thermal corrections to vacuum polarization generating a nonzero and significantly large longitudinal component of the polarization tensor which could not be ignored. Thermal contribution to electrical permittivity (Eq. 8a) and the magnetic permeability $\mu(K)$ (Eq. 8b) are used to evaluate thermal contribution to the propagation speed of electromagnetic waves and other relevant parameters in the early universe and it is given by,

$$v_{prop} = \sqrt{\frac{1}{\varepsilon(K) \mu(K)}} \tag{9a}$$

Which correspond to the refractive index $r_i$ of the medium as:

$$r_i = \frac{c}{v} = \sqrt{\frac{\varepsilon(K) \mu(K)}{\varepsilon_0(K) \mu_0(K)}} \tag{9b}$$

Longitudinal component of the vacuum polarization tensor at finite temperature can be used to determine the phase of the medium indicating overall properties of the medium. Medium properties are changing with temperature due to modified electromagnetic couplings (2c). $K_L$ is evaluated by taking $\omega = k_0 = 0$ and p very small, the Debye shielding length $\lambda_D$ of such a medium is then given by the inverse of $K_L$

$$\lambda_D = 1/K_L \tag{10a}$$

and the corresponding $\omega_D$ is

$$\omega_D = \frac{2\pi v_{prop}}{\lambda_D} = 2\pi K_L v_{prop} \qquad (10b)$$

Satisfying the relation

$$f_D \lambda_D = v_D \qquad (10c)$$

The plasma frequency $\omega_P$ is also related to $\omega_T$ from $\Pi_T(\omega = |k|)$ as

$$\omega_D = \frac{2\pi v_D}{\lambda_D} \qquad (11)$$

The longitudinal and transverse components of the photon frequency ω and the momentum k can be evaluated at different temperatures for given values of ω and k. However their effective values are determined as

$$k = (k_L^2 + k_T^2)^{1/2} \qquad (12a)$$

Such that

$$\omega = (\omega_L^2 + \omega_T^2)^{1/2} \qquad (12b)$$

The plasma frequency is defined as $\omega_P^2 = \omega_T^2$ as $\omega_L^2 = 0$ and the Debye shielding length is obtained from the longitudinal component of wavelength using equation (10b). These results can easily be generalized to different situations using the initial values of $\Pi_T$ and $\Pi_L$.

1. **RESULTS and DISCUSSIONS**

Physically measureable values of QED parameters in the extremely hot universe, right after the big bang, becomes temperature dependent due to its interaction with the hot fermions of the universe. Massless photons acquire dynamically generated mass in such a medium and the QED coupling depends on temperature. We discuss the propagation of monochromatic light in this extremely hot universe at $T \geq 10^{10}$ K in the presence of hot fermions in the medium. Temperature is measured in units of electron mass (0.511 MeV), which corresponds to the temperature T ($\sim 10^{10}$ Kelvin). Natural system of units is used throughout this study such that the speed of light c =1 and mass, energy, temperature and momentum are all expressed in units of electron mass. Boltzmann constant $k_B$ is set equal to one in vacuum whereas the electromagnetic properties such as electric permittivity, magnetic permeability and dielectric constant are normalized to unity in this system of units, for convenience. In the real-time formalism, thermal contribution appears as additive contribution to vacuum values, especially at the one loop level. Thermal corrections appear as temperature dependent additive terms and can be analyzed independent of the corresponding vacuum values. Fig.1 gives the plot of $k_L^2$ and $\omega_T^2$ as a function of temperature. It is clearly seen that for T below the electron mass, $\omega = k_0 = 0$. For larger T, these two functions develop a quadratic dependence on temperature and start to decouple though the relatively slower growth in the transverse frequency as compared to the relatively faster growth in longitudinal momentum k. The larger value of the longitudinal component of propagation vector $k_L$ indicate the dominant interaction of electromagnetic signals in the medium as compared to the growth in transverse frequency $\omega_T$ showing more bending in the small universe with slower increase in transverse oscillation and gives an explanation of signal trapping in the smaller and hotter universe.

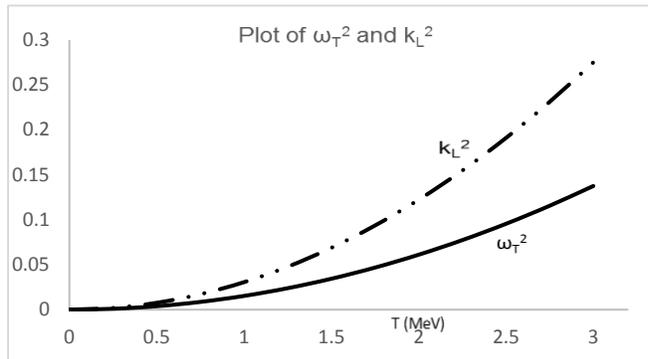

*Fig.1: Temperature dependence of the longitudinal component of the propagation vector (square) and the transverse frequency (square) are plotted showing that the Debye shielding length is greater than the plasma frequency.*

We plot thermal contribution just around the nucleosynthesis temperature and below the neutrino decoupling. T=3 m correspond to temperature around 1.5 MeV which is just close to the nucleosynthesis temperature. So pure QED plasma is not expected at T higher than neutrino decoupling. This temperature is well below the neutrino decoupling. This range is chosen to see how the effect of temperature becomes significant and the electromagnetically interacting medium may exist in plasma phase. Temperature has to be considered sufficiently below neutrino decoupling to assure the insignificance of thermal contribution of electroweak interaction as there should not be enough neutrinos in the medium. On the other hand, it is obvious from Fig. 2 that the transverse component of propagation vector is not much affected by temperature as it just depends on the $b(m\beta)$ and is ignorable as compared to quadratic function $c(m\beta)$.

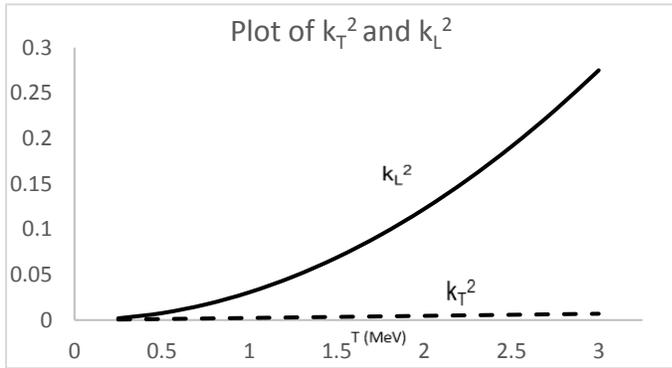

Fig. 2: Comparing the plots of $k_L$ (square) and $k_T$ (square), it can be clearly seen that propagation vector in the transverse direction is independent of temperature. Temperature increases the inverse of Debye length ($1/k_L$) but transverse motion will not be affected.

Moreover, the transversality of electromagnetic signals requires $k_L^2=0$ at temperature below the electron mass due to the absence of coupling with the medium which is not ignorable due to the radiative corrections at $T>10^9$ K. Negligible change in $k_T^2$ gives the assurance that the transverse speed remains almost the same thus the $k_L^2$ has quadratic dependence on T which indirectly affect the propagation velocity by mainly bending and then slowing it down inside the plasma, which is a king of trapping in plasma. It seems to make perfect sense as the smaller volumes at higher densities and higher temperatures have more bending than the lower temperatures where the size of the universe was larger due to the expansion. These plots show how the plasma screening was dissolved as the nucleosynthesis was almost complete around temperature ($T \approx m$) when electrons and photons were not shielding each other. Neutrinos were needed to initiate nucleosynthesis and their concentration was reduced during nucleosynthesis.

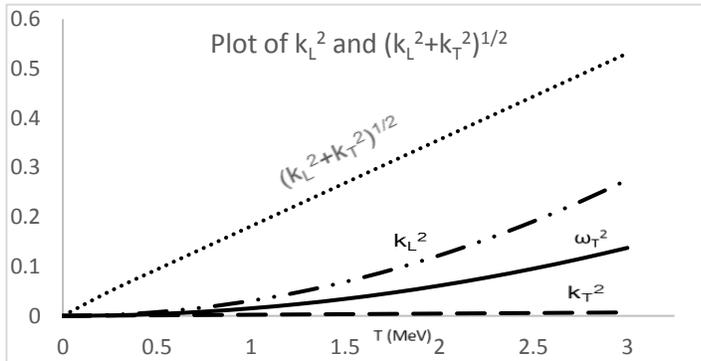

Fig. 3: Plots of k, the magnitude of the propagation vector and the longitudinal propagation $k_L$ (square) and transverse frequency $\omega_T$ (square). k linearly depends on T whereas all other parameters have quadratic dependence on T.

When the plasma screening is large enough, it traps the electromagnetic signals and the measurable value of the speed of electromagnetic signals is not a constant quantity any more. It starts to depend on temperature as well. Propagation with the transverse frequency makes it phase velocity which is responsible for oscillation of electromagnetic waves within the Debye sphere of length proportional to $1/k_L$. To clearly identify the unique behavior of electromagnetic signals in this range by comparison, we plot the magnitude of wave vector k along with

$k_L^2$ and $\omega_T^2$ in Fig.3. It is obvious that k is directly proportional to T and is contributed both by the longitudinal and transverse components.

On the other hand the longitudinal component of the wavenumber k correspond to the inverse of Debye length, increasing $k_L$ indicates the decrease in the Debye shielding length with temperature. Quadratic increase in $k_L^2$ correspond to inverse proportionality of Debye length with temperature.

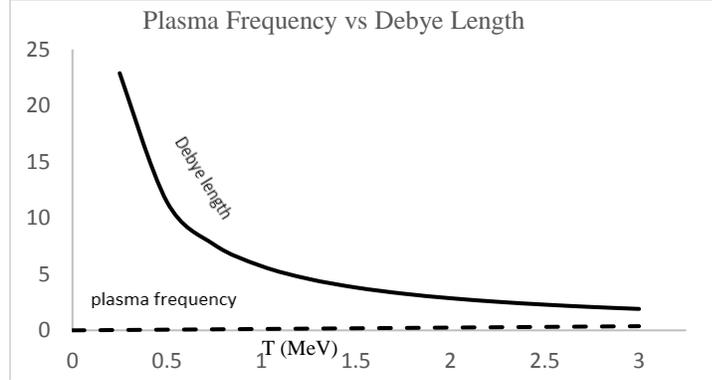

*Fig. 4: Debye length ($1/k_L$) and the Plasma frequency are plotted as a function of temperature showing that the screening length increases with T. Debye length is significantly large around $T \approx m$ indicating the presence of QED plasma.*

Fig. 4 gives a plot of Debye length and plasma frequency which correspond to the plots of Eq. (10a) and (11). Appearance of a small longitudinal component around $T \approx m$ electron mass shows that the transverse component of k is significant at T< 2MeV also. Then the transverse contribution is less significant as compared to the longitudinal component as it remains unchanged (practically) with temperature and the corresponding longitudinal component grows. It has been previously shown that the temperature dependence is a functions of Masood's $a_i$ functions only near the nucleosynthesis temperature. Otherwise, the quadratic dependence on temperature is proven at lower T as well as the larger values of T. Refs. [5, 6, 12, 13] discuss this in detail. A detailed investigation of QED behavior can also be found in literature [21-23]. Fig. 5 gives a plot of QED effect at larger temperatures. If we ignore any other effect to check the behavior of QED at higher than the neutrino decoupling temperatures, a phase transition occurs at around 16 MeV and an unusual behavior of trapping of light is shown above those temperature. It also indicates an extremely large coupling of QED. This unusual behavior automatically suggest the presence of some other types of interaction. Thus the electromagnetic properties of such a medium are tremendously changed and a weak interaction contribution is needed to be incorporated.

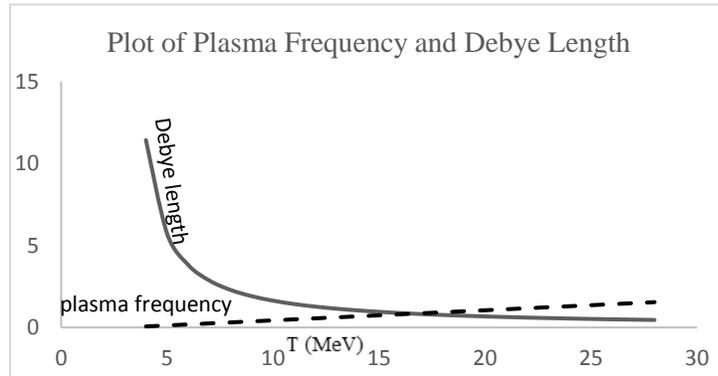

*Figure 5: Debye length ($1/k_L$) and the Plasma frequency are plotted as a function of temperature showing that the screening length increases with T. When temperature rises around 16 times the mass of electron which is around 8MeV, a phase transition takes place and the plasma frequency becomes greater than the*

Eqns. (7-11) in section 3 help to figure out the properties of the QED medium at high temperatures where the plasma effects can be seen. However, the photon propagation through the medium could be studied for other relevant ranges of the photon frequency and wavenumber also.

For small values of k, the vacuum relation $v_{prop} \approx c = 1$ is still valid and $v_{prop} \approx 1 \approx r_i$ is reproduced or such a medium can be considered as a transparent medium. Small k value indicates that the wave number is negligible and the photon frequency is very high. It means that the medium of the early universe was transparent for the extremely high energy radiation which is a naturally acceptable result. We expect the extremely high energy radiation in the early universe and that is what we should expect in the early universe. Understanding of propagation of light will help us to understand the inflation and generation of anisotropy

[17-23] in the early universe. A similar type of effect has already been seen in QCD using the perturbative calculations of QCD in the real-time formalism [24-28]. Effect of temperature on the interaction of neutrino with the photon has already been calculated [15-20].

Just to summarize, this paper gives a unique approach to study the universe for a short interval of time between nucleosynthesis and neutrino decoupling. Renormalization techniques of QED are used to prove that the universe behaved as a relativistic plasma during the time of nucleosynthesis and before the neutrino decoupled. These limits are determined from the validity of scheme of calculations. Thermal effects are ignorable before nucleosynthesis as we did not have enough electrons at extremely high temperature to give significant thermal contribution to the QED coupling whereas the electroweak interaction is not ignorable after neutrino decoupling so the QED renormalization scheme will not be fully describing the system anymore.

## REFERENCES AND FOOTNOTES